

\font\titlefont = cmr10 scaled\magstep 4
\font\sectionfont = cmr10
\font\littlefont = cmr5
\font\eightrm = cmr8

\def\ss{\scriptstyle}
\def\sss{\scriptscriptstyle}

\magnification = 1200

\global\baselineskip = 1.2\baselineskip
\global\parskip = 4pt plus 0.3pt
\global\abovedisplayskip = 18pt plus3pt minus9pt
\global\belowdisplayskip = 18pt plus3pt minus9pt
\global\abovedisplayshortskip = 6pt plus3pt
\global\belowdisplayshortskip = 6pt plus3pt


\def\endignore{}
\def\ignore #1\endignore{}

\newcount\dflag
\dflag = 0


\def\monthname{\ifcase\month
\or Jan \or Feb \or Mar \or Apr \or May \or June%
\or July \or Aug \or Sept \or Oct \or Nov \or Dec
\fi}

\def\timestring{{\count0 = \time%
\divide\count0 by 60%
\count2 = \count0
\count4 = \time%
\multiply\count0 by 60%
\advance\count4 by -\count0
\ifnum\count4 < 10 \toks1 = {0}
\else \toks1 = {} \fi%
\ifnum\count2 < 12 \toks0 = {a.m.}
\else \toks0 = {p.m.}
\advance\count2 by -12%
\fi%
\ifnum\count2 = 0 \count2 = 12 \fi
\number\count2 : \the\toks1 \number\count4%
\thinspace \the\toks0}}

\def\today{\ifcase\month\or January\or February\or March\or
 April\or May\or June\or July\or August\or September\or
 October\or November\or December\fi \space\number\day, \number\year}



\def\endtitle{}
\def\title#1\endtitle{\vskip.5in\titlefont
\global\baselineskip = 2\baselineskip
#1\vskip.4in
\baselineskip = 0.5\baselineskip\rm}

\def\endauthors{}
\def\authors#1\endauthors{#1}

\def\endabstract{}
\def\abstract#1\endabstract{\vskip .3in%
\centerline{\sectionfont\bf Abstract}%
\vskip .1in
\noindent#1}

\newcount\nsection
\newcount\nsubsection

\def\section#1{\global\advance\nsection by 1
\nsubsection=0
\bigskip\noindent\centerline{\sectionfont \bf \number\nsection.\ #1}
\bigskip\rm\nobreak}

\def\subsection#1{\global\advance\nsubsection by 1
\bigskip\noindent\sectionfont \sl \number\nsection.\number\nsubsection)\
#1\bigskip\rm\nobreak}

\def\topic#1{{\medskip\noindent $\bullet$ \it #1:}}
\def\endtopic{\medskip}

\def\appendix#1#2{\bigskip\noindent%
\centerline{\sectionfont \bf Appendix #1.\ #2}
\bigskip\rm\nobreak}


\newcount\nref
\global\nref = 1

\def\ref#1#2{\xdef #1{[\number\nref]}
\ifnum\nref = 1\global\xdef\therefs{\noindent[\number\nref] #2\ }
\else
\global\xdef\oldrefs{\therefs}
\global\xdef\therefs{\oldrefs\vskip.1in\noindent[\number\nref] #2\ }%
\fi%
\global\advance\nref by 1
}

\def\listrefs{\vfill\eject\section{References}\therefs}


\newcount\nfoot
\global\nfoot = 1

\def\foot#1#2{\xdef #1{(\number\nfoot)}
\footnote{${}^{\number\nfoot}$}{\eightrm #2}
\global\advance\nfoot by 1
}


\newcount\nfig
\global\nfig = 1

\def\fig#1{\xdef #1{(\number\nfig)}
\global\advance\nfig by 1
}


\newcount\cflag
\newcount\nequation
\global\nequation = 1
\def\eqlabel{(1)}

\def\nexteqno{\ifnum\cflag = 0
\global\advance\nequation by 1
\fi
\global\cflag = 0
\xdef\eqlabel{(\number\nequation)}}

\def\lasteqno{\global\advance\nequation by -1
\xdef\eqlabel{(\number\nequation)}}

\def\label#1{\xdef #1{(\number\nequation)}
\ifnum\dflag = 1
{\escapechar = -1
\xdef\draftname{\littlefont\string#1}}
\fi}

\def\clabel#1#2{\xdef\eqlabel{(\number\nequation #2)}
\global\cflag = 1
\xdef #1{\eqlabel}
\ifnum\dflag = 1
{\escapechar = -1
\xdef\draftname{\string#1}}
\fi}

\def\cclabel#1#2{\xdef\eqlabel{#2)}
\global\cflag = 1
\xdef #1{\eqlabel}
\ifnum\dflag = 1
{\escapechar = -1
\xdef\draftname{\string#1}}
\fi}


\def\eeq{}

\def\eqnn #1\eeq{$$ #1 $$}

\def\eq #1\eeq{\xdef\draftname{\ }
$$ #1
\eqno{\eqlabel \rlap{\ \draftname}} $$
\nexteqno}




\def\eeol{& \eqlabel \rlap{\ \draftname}
\nexteqno
\xdef\draftname{\ }}

\def\eolnn{\cr
\global\cflag = 0
\xdef\draftname{\ }}


\def\eqa #1\eeq{\xdef\draftname{\ }
$$ \eqalignno{ #1 } $$
\global\cflag = 0}


\def\etal{{\it et.al.\/}}

\def\cf{{\it c.f.\/}}


\def\jetpl#1#2#3#4#5#6{{\it Pis'ma Zh.~Eksp.~Teor.~Fiz.} {\bf #1} (19#2) #3
[{\it JETP Lett.} {\bf #4} (19#5) #6]}

\def\plb#1#2#3{{\it Phys.~Lett.} {\bf #1B} (19#2) #3}

\def\prd#1#2#3{{\it Phys.~Rev.} {\bf D#1} (19#2) #3}

\def\prep#1#2#3{{\it Phys.~Rep.} {\bf C#1} (19#2) #3}
\def\prl#1#2#3{{\it Phys.~Rev.~Lett.} {\bf #1} (19#2) #3}
\def\rmp#1#2#3{{\it Rev.~Mod.~Phys.} {\bf #1} (19#2) #3}


\global\nulldelimiterspace = 0pt



\def\frac#1#2{{{#1} \over {#2}}\,}  
\def\hf{{1\over 2}}



\def\Dsl{\hbox{/\kern-.6700em\it D}} 
\def\dsl{\hbox{/\kern-.5300em$\partial$}}
\def\pxpsl{\hbox{/\kern-.5600em$p$}}
\def\ssl{\hbox{/\kern-.5300em$s$}}
\def\epssl{\hbox{/\kern-.5100em$\epsilon$}}
\def\delsl{\hbox{/\kern-.6300em$\nabla$}}
\def\lxpsl{\hbox{/\kern-.4300em$l$}}
\def\elxpsl{\hbox{/\kern-.4500em$\ell$}}
\def\kxpsl{\hbox{/\kern-.5100em$k$}}
\def\qxpsl{\hbox{/\kern-.5000em$q$}}
\def\sla#1{\raise.15ex\hbox{$/$}\kern-.57em #1}
\def\Pl{\gamma_{\sss L}}

\def\pwr#1{\cdot 10^{#1}}



\def\roughly#1{\mathrel{\raise.3ex
    \hbox{$#1$\kern-.75em\lower1ex\hbox{$\sim$}}}}
\def\lsim{\roughly<}
\def\gsim{\roughly>}

\def\tw#1{\tilde{#1}}
\def\ol#1{\overline{#1}}



\def\bfg{{\bf g}}

\def\bfm{{\bf m}}



\def\Scl{{\cal L}}

\def\Scn{{\cal N}}

\def\Scs{{\cal S}}

\def\Scw{{\cal W}}
\def\Scx{{\cal X}}




\def\avg#1{\langle #1 \rangle}

\def\Avg#1{\left\langle #1 \right\rangle}



\def\cc{{\rm c.c.}}

\def\eV{{\rm \ eV}}
\def\keV{{\rm \ keV}}
\def\MeV{{\rm \ MeV}}
\def\GeV{{\rm \ GeV}}


\def\msbar{$\overline{\hbox{MS}}$}
\def\SNt#1#2{$#1 \times 10^{#2}$}
\def\SN#1#2{#1 \times 10^{#2}}
\def\ngb{Nambu-Goldstone boson}
\def\lft{{\sss L}}
\def\rht{{\sss R}}
\def\mee{m_{ee}}
\def\memu{m_{e\mu}}

\def\pf{p_{\sss F}}
\def\EF{E_{\sss F}}

\def\GF{G_{\sss F}}
\def\bb{\beta\beta}
\def\bbtn{\bb_{2\nu}}

\def\bbzn{\bb_{0\nu}}

\def\Nd{$^{150}$Nd}
\def\Mo{$^{100}$Mo}
\def\Se{$^{82}$Se}
\def\Nd{$^{150}$Nd}

\def\Ge{$^{76}$Ge}

\def\ol#1{\overline{#1}}
\def\mnu#1{m_{\nu_#1}}
\def\uiui{\hbox{U(1)$_e \times$U(1)$_\mu$}}
\def\stth{\sin^2 (2\theta_{e\mu})}


\rightline{McGill-93/10}
\rightline{hep-ph/9306252}
\rightline{May 1993}
\vskip .2in

\title
\centerline{ A Common Explanation for the }
\centerline{ Atmospheric, Solar-Neutrino and }
\centerline{ Double Beta Decay Anomalies }
\endtitle

\authors
\centerline{C.P. Burgess and Oscar F. Hern\'andez\footnote{}{email:
cliff@physics.mcgill.ca; oscarh@physics.mcgill.ca}}
\vskip .15in
\centerline{\it Physics Department, McGill University}
\centerline{\it 3600 University St., Montr\'eal, Qu\'ebec, CANADA, H3A 2T8.}
\vskip .1in
\endauthors

\abstract
Acker \etal\ have recently proposed an economical solution to the solar and
atmospheric neutrino deficits, in which both are explained by large-angle
$\nu_e - \nu_\mu$ oscillations, supplemented by $\nu_e$ decays. We show how to
embed their phenomenological model into an electroweak framework in which
global electron and muon numbers, (U(1)$_e \times$U(1)$_\mu$), spontaneously
break at a scale of 1 keV. Despite such a low scale, our model is technically
natural. The naturalness requirement, together with nucleosynthesis
constraints,
point to the existence of relatively light, largely sterile neutrinos with
masses in the MeV range. We find a number of potentially interesting
experimental  implications of these models, one of which is an explanation of
the excess events that have been found near the endpoints in the double beta
decay of several elements. One formulation of our model involves a novel
realization of supersymmetry, for which the new light particles and their
superpartners are split by very small amounts in comparison with the weak
scale.
\endabstract


\vfill\eject
\section{ Introduction }

The various experimental neutrino anomalies that have come to light over the
past years fall into two categories according to whether they can be explained
simply by a particular pattern of neutrino masses and mixings, or whether they
require in addition the inclusion of new, light scalars at very low energies.
The solar neutrino problem
\ref\solarexp{K.S. Hirata \etal, \prl{66}{91}{9}; \prl{65}{90}{1301};
K. Nakamura, {\it Nucl. Phys.} (Proc. Suppl.) {\bf B31} (1993);
P. Anselmann \etal, \plb{285}{92}{376};
A.I. Abazov \etal, \prl{67}{91}{3332};
V.N. Gavrin, presented at ICHEP XXVI, Dallas, Aug. 1992;
K. Lande \etal, in Proc. XXVth Int. Conf. on High Energy Physics,
ed. by K.K. Phua and Y. Yamaguchi, (World Scientific, Singapore, 1991).}
\ref\solarth{J.N. Bahcall and M.H. Pinsonneault, \rmp{64}{92}{885};
S. Turck-Chi\`eze \etal, {\it Astrophys. J.} {\bf 335} (1988) 415.}
\solarexp, \solarth\ and the atmospheric neutrino deficit
\ref\atmospheric{K.S. Hirata \etal, \plb{280}{92}{146}; \plb{205}{88}{416};
D.W. Casper \etal, \prl{66}{91}{2561};
R. Becker-Szendy \etal, preprint BUHEP-91-24;
M. Goodman, presented to DPF-92 meeting, Fermilab, Nov. 1992.}
\atmospheric\ are in the first of these categories, while the observed excess
of
electrons in the double-beta ($\bb$) decay of \Ge, \Mo, \Se, and \Nd\
\ref\dbeta{M. Moe, M. Nelson, M. Vient and S. Elliot, preprint UCI-NEUTRINO
92-1;
M. Moe, {\it Nucl. Phys.} (Proc. Suppl.) {\bf B31} (1993).}
\dbeta\ fall into the latter group. This basic difference makes it difficult to
understand all three as being different manifestations of the same type of new
physics.

Recently Acker \etal\
\ref\acker{A. Acker, A. Joshipura and S. Pakvasa, \plb{285}{92}{371};
A. Acker, J.G. Learned, S. Pakvasa and T. Weiler, \plb{298}{93}{149}.}
\acker\ have proposed an economical model within which both the solar and
atmospheric neutrino deficits are simultaneously explained using only electron
and muon neutrinos. In their approach, the atmospheric $\nu_\mu/\nu_e$ ratio is
depleted by $\nu_e - \nu_\mu$ oscillations with $\Delta m^2 \simeq 10^{-2} \,
\eV^2$, and $\stth \simeq 0.8$. This region of parameter space is unconstrained
by the IMB limits on upward-going muons
\ref\imb{R. Becker-Szendy \etal, \prl{69}{92}{1010}.}
\imb, once matter-effects are included for the propagation of $\nu_e$'s
through the earth
\ref\priv{S. Pakvasa, private communication.}
\priv. Besides being depleted by these same oscillations, solar neutrinos are
further suppressed in this model by neutrino decays {\it en-route} from the
sun: $\nu_h \to \nu_\ell + \varphi$. In these decays $\varphi$ is a scalar
particle -- perhaps a majoron -- and the heavier mass eigenstate, $\nu_h$, is
dominantly $\nu_e$. It is this additional decay that permits a solar-flux
signal
at Homestake that is less than half of what the standard solar model predicts,
a
suppression that would be impossible with only large-angle two-flavour
oscillations.

The one experimental drawback of the model is the energy-dependence that it
predicts for the solar-neutrino reduction, since neutrino decays
preferentially suppress the flux of lower-energy {\it p-p}-cycle neutrinos that
are seen in the Ga experiments in comparison with the higher-energy ${}^8$B
neutrinos which dominate the signal in Cl
\ref\decays{Other references to neutrino decay solutions to the solar neutrino
problem are: \hfil\break
J.N. Bahcall, N. Cabibbo and A. Yahill, \prl{28}{72}{316};
 S. Pakvasa and K. Tennakone, \prl{28}{72}{1415};
J.N. Bahcall \etal, \plb{181}{86}{369};
Z.G. Berezhiani and M.I. Vysotsky, \plb{199}{87}{281};
J.A. Frieman, H.E. Haber and K. Freese, \plb{200}{88}{115};
A. Acker, S. Pakvasa and J. Pantaleone, \prd{43}{91}{1754};
Z.G. Berezhiani, G. Fiorentini, M. Moretti and A. Rossi,
\jetpl{55}{92}{159}{55}{92}{151}.}
\decays. This suppression arises because the perceived lifetime of the more
energetic particles is longer (due to time dilation) in the rest frame of the
solar system. This prediction is ruled out up to the 95\% C.L.~by the
observations, which indicate that the Homestake solar-neutrino suppression is
larger than that seen in GALLEX.

The advantage of this picture, on the other hand, is that the use of neutrino
decays into scalars casts the solution to the atmospheric and solar problems in
a way which is very similar to the interpretation of the excess $\bb$ events as
the signal of scalar particle emission
\ref\jim{C.P. Burgess and J.M. Cline, \plb{298}{93}{141}, and preprint
McGill-93/02, TPI-MINN-93/28-T, UMN-TH-1204-93 (unpublished).}
\ref\smirnov{Z.G. Berezhiani, A.Yu. Smirnov and J.W.F. Valle,
\plb{291}{92}{99}.}
\jim, \smirnov. This raises the possibility of accounting for all three of
these
experimental anomalies using the same kind of new physics. With this in mind,
we
choose here to reserve judgement on the potential discrepancy with the relative
size of the GALLEX and Homestake experiments, in order to further explore
whether the $\bb$ anomaly can also be described. In any case, given the
difficulty of these experiments, with the potential for unknown systematic
errors, we regard this discrepancy as being merely preliminary.

Our second goal is to come to grips with the fine-tuning problem that plagues
all such models which involve very light scalars that are relatively
strongly-coupled to neutrinos \jim. In some models this fine tuning arises
because a scalar mass must be directly tuned to be extremely small (tens of keV
or less). Alternatively, if the light scalar is a (pseudo-) \ngb, its coupling
to
neutrinos is typically of order $g_\nu \simeq m_\nu/f$, where $m_\nu$ is a
relevant neutrino mass and $f$ is the scale of symmetry breaking. The
requirement that $g_\nu$ be large enough to reproduce the anomalous experiments
together with the upper limit on $m_\nu$ typically implies a small
symmetry-breaking scale, $f$ (also tens of keV or less). In either case a very
small scalar symmetry-breaking scale is required, whose stability under
renormalization must be established.

We address these issues by embedding the Acker \etal\ theory into a
renormalizable model of physics at the weak scale and below. Within
this framework we are able to find the conditions under which the
model can solve the atomospheric and solar neutrino problems, as well
as the $\bb$ excess, in a technically natural way. By technical
naturalness we mean that renormalization corrections to small
dimensionless parameters are automatically of the same size as, or
smaller than the tiny parameter itself. This condition is satisfied by
all known heirarchies of scale in physics, and so is the bare minimum
that might be asked of a reasonable theory. It is a {\it weaker}
condition than 't Hooft naturalness, which would also demand that
additional symmetries arise as the value of the small parameter goes
to zero.

We find that the conditions for naturalness are very restrictive, particularly
when taken together with what is needed to describe the $\bb$ decay excess and
nucleosynthesis constraints. As we show in detail below, a renormalizable
description of the new physics requires the existence of new degrees of
freedom, which we take to be sterile neutrinos. Primordial nucleosynthesis and
$\bb$ decay argue that these new states should have masses larger than of order
MeV in scale. This constraint, which tends to drive the new physics to higher
energy scales, contrasts with the naturalness constraint which prefers the new
particles to be lighter than 1~MeV. These bounds work against each other, and
are contradictory in the generic case. We find that all constraints {\it
can} be accomodated, however, if the physics of the weak scale should be
supersymmetric. Supersymmetry (SUSY) helps because it weakens the strength of
the naturalness constraints, permitting them to coexist with the cosmological
and
$\bb$ requirements.

We present a renormalizable realization of both the supersymmetric and
nonsupersymmetric scearios here, and find that, for both of these, the
combination of all of the bounds points to the existence of sterile neutrinos
in
both the few MeV and the several-hundred MeV ranges. The several-MeV neutrino
is
predominantly mixed with $\nu_\mu$, and the several-hundred-MeV neutrino mixes
most significantly with $\nu_e$. Both types of mixings may be amenable to
detection through closer scrutiny of the $\pi \to \mu \nu$ decay spectrum, as
well as in weak-universality measurements of the electron charged-current
interaction strength.

\section{ The Atmospheric and Solar Neutrino Deficits }

We start by reproducing the model of Acker \etal\ \acker\ as the
following low energy effective theory. In addition to the
standard-model (SM) particles and gauge symmetries we require the
conservation of a global \uiui\ electron- and muon-number symmetry,
which will eventually be spontaneously broken by two additional
electroweak-singlet scalars which transform under the lepton
symmetries in the following way:
\label\scalarcharges
\eq
\phi_e \sim (-2,0) \qquad \phi_{e\mu} \sim (-1,-1)
\eeq
Two scalars are required here in order to obtain neutrino decays, since a
single scalar produces Yukawa couplings that are automatically diagonal in a
basis of neutrino mass eigenstates.

Under this symmetry only the following SM particles carry nontrivial charge
\label\charges
\eqa \psi_{e\lft}=\left(\matrix{\nu_e\cr e_\lft \cr}\right) & \sim (1,0)
    \qquad e_\rht \sim (1,0) \eolnn
     \psi_{\mu\lft}=\left(\matrix{\nu_e\cr e_\lft \cr}\right) & \sim (0,1)
     \qquad \mu_\rht \sim (0,1) \eeol \eeq
With these charge assignments, and denoting the SM Higgs by $H$, the lowest
dimension terms that lead to neutrino masses and mixing arise at dimension six:
\label\effdimsix
\eq
{1\over 2\Lambda^2}\left(\lambda_{e}\psi_e\psi_e H H \phi_e
+ 2 \lambda_{\mu}\psi_e\psi_\mu H H \phi_{e\mu} \right) .
\eeq
This effective description contains all the information that is relevant for
atmospheric- and solar-neutrino physics,
(and as we will see in section 3, the double-beta decay anomaly as well).

Replacing $H$, $\phi_e$ and $\phi_{e\mu}$ by their respective {\it v.e.v.}'s,
$v=$174 GeV, $w_e$, and $w_\mu$, gives rise to the following mass matrix and
Yukawa couplings:
\label\lowmass
\eq
\bfm = \pmatrix{	\mee 	& \memu \cr	\memu & 0	\cr }, \qquad
\bfg = \pmatrix{	\mee/w_e 	& \memu/w_\mu \cr	\memu/w_\mu & 0	\cr }.
\eeq
where $\mee = \lambda_e v^2 w_e/\Lambda^2$ and $\memu = \lambda_\mu v^2
w_\mu/\Lambda^2$. The mass eigenstates and their masses work out to be:
\label\loweigen
\eq \eqalign{
\nu_h  = \left(\matrix{c \cr s \cr}\right) &
\qquad \nu_\ell = \left(\matrix{-s \cr c \cr}\right) \cr
m_h = \hf \left( \sqrt{\mee^2+4\memu^2} + \mee \right) &\qquad
m_\ell = \hf \left( \sqrt{\mee^2+4\memu^2} - \mee \right), \cr}
\eeq
where we define $c\equiv\cos\theta_{e\mu}$ and $s\equiv\sin\theta_{e\mu}$ and
\label\thetadef
\eq
\stth = {4 m_\ell/m_h \over (1 + m_\ell/m_h)^2} .
\eeq
The choice $\sin^2(2\theta_{e\mu})\approx 0.8$ corresponds to $\mee \approx
\pm\memu$, and $\Delta m^2 \sim 10^{-2}\, \eV^2$
then gives $m_h \sim 0.1 \, \eV$ and $m_\ell \sim 0.04 \, \eV$.
(Here $\Delta m^2 \equiv m_h^2 -m_l^2$.)

The spectrum of light scalars contains two massive states with masses $\lsim
w_i$, as well as two massless  \ngb s\foot\anomaly{Any masses these states may
obtain due to the electroweak anomalies in the global symmetry group are
completely negligible for our purposes.} which we take to be the (appropriately
normalized) phases of the two scalar fields: $\chi_e$ and $\chi_{e\mu}$.
Anticipating the result $w_i \sim 1 \, \keV$ that is required to produce the
desired neutrino lifetime implies that only the massless states are relevant to
neutrino decays. Their low-energy neutrino couplings are:
\label\goldcoupling
\eq  \Scl_{\chi \nu\nu} =
{-i \over 4\sqrt2} \left[ (\ol{\nu}_{e} \gamma^\alpha \gamma_5 \nu_{e})
{ \partial_\alpha \chi_e \over w_e} + (\ol{\nu}_{\mu} \gamma^\alpha \gamma_5
\nu_{\mu}) \left(2{\partial_\alpha \chi_{e\mu} \over w_\mu} -
 {\partial_\alpha \chi_{e} \over w_e} \right)  \right],
\eeq
which produces the following total rate for $\nu_h$ decay, in a frame for which
$\nu_h$ has energy $E$:
\label\decay
\eq
\Gamma = {\stth \over 128 \pi E} \;{ ( \Delta m^2 )^3 \over m_h^2} \;
\left( {1 \over w_e^2 } + {1 \over w_\mu^2} \right). \eeq

It is easy to check that this effective lagrangian solves the atmospheric and
solar neutrino problems in the way envisioned by Acker \etal, as long as the
{\it v.e.v.} of the scalars is about 1 keV,
which with the conditions below eq.~\thetadef\ implies
 the coupling $m_{ei}/w_i =
\lambda_i v^2/\Lambda^2 \sim 10^{-4}$.

\section{ Double Beta Decay }

The remarkable fact is that these same values that resolve the solar- and
atmospheric-neutrino problems are also just what is needed to account for the
excess high-energy electrons near the end point of the double beta
decay spectrum. We establish this point in the present section.

The differential decay rate for double-beta decay of a nucleus of charge $Z$
and mass number $A$ accompanied by the emission of a light scalar particle is
given by \jim:
\label\bbdecayrate
\eq {d \Gamma \over d\varepsilon_1 d\varepsilon_2} = { \GF^4 \over 32 \pi^5}
| \Scw |^2 (Q - \varepsilon_1 - \varepsilon_2) [ \varepsilon_1 p_1
F(\varepsilon_1) ] [\varepsilon_2 p_2 F(\varepsilon_2) ]. \eeq
where $\GF$ is Fermi's constant; $Q = M(Z,A) - M(Z+2,A) - 2m_e$ is the kinetic
energy --- typically several MeV --- that is available to the final-state
leptons; $\varepsilon_i$ and $p_i$ are the energy and momentum of each of the
final two electrons; and $F(\varepsilon)$ is the Fermi function which describes
the distortion of the electron spectrum due to the nuclear charge.

$\Scw$ represents the following integral:
\label\wdef
\eq  \Scw \equiv \sum_{ij} V_{ei} V_{ej} \int {d^4q \over (2 \pi)^4}
\; \left[ { \mnu{i} a_{ij} \mnu{j} - q^2 b_{ij} \over (q^2 + \mnu{i}^2
- i\epsilon) (q^2 + \mnu{j}^2 - i\epsilon) } \right] w(q^0,q^2). \eeq
The sum here is over all neutrino species, with $V_{ei}$ representing the
strength (normalized so that $V_{ei} = \delta_{ei}$ in the standard model) of
the
$e - \nu_i$ charged-current interaction. $a_{ij}$ and $b_{ij}$ are
coupling matrices defined by the following Yukawa interaction:
$\Scl = - \hf\,
\ol{\nu}_i \,(b_{ij} \phi + a_{ij} \phi^*) \Pl \nu_j + \cc$. $w$ represents a
particular combination of Lorentz-invariant form factors, as defined in
Ref.~\jim, which describe the nuclear matrix element of two hadronic charged
currents.

For the purposes of analyzing our model, we make the following three
simplifying
assumptions:
\topic{1} We neglect the electron mass, $m_e \simeq 0$, in performing all phase
space integrals,
\topic{2} We neglect the Coulomb-distortion factor, $F(\varepsilon) \simeq 1$,
and
\topic{3} We parameterize the nuclear form factors by approximating them by
step functions in energy and momentum: $w \simeq w_0 \, \Theta(q^0 - \EF) \,
\Theta(|\vec{q}| - \pf)$. Here $\pf$ represents the nucleon Fermi momentum
and $\EF = \pf^2/2m_{\sss N}$ is the corresponding Fermi energy. We fix the
normalization, $w_0 \simeq 4 \, \MeV^{-1}$, by comparing to the observed
$\bbtn$
half lives, and we fix $\pf \simeq 60 \, \MeV$ (and so $\EF \simeq 2$ MeV)
by requiring that the present
upper limit on the half life for neutrinoless decay
\ref\bbznbound{A. Piepke, presented at ICHEP XXVI, Dallas, Aug. 1992;
D. Caldwell \etal, {\it Nucl. Phys.} (Proc. Suppl.) {\bf B13} 1990 547.}
\bbznbound, $T_\hf(\bbzn) > 2\pwr{24}$ yr, imply the corresponding upper bound
on the $\nu_e$ majorona mass: $\mnu{e} < 1 \, \eV$
\ref\bbzntheory{T. Tomoda, {\it Rep. Prog. Phys.} {\bf 54} (1991) 53.}
\bbzntheory. These approximations
produce similar results to the more detailed calculations of Ref.~\jim, and
are sufficiently accurate to determine whether our model can account for the
observed anomaly.
\endtopic

With these choices, and neglecting the neutrino masses in comparison with the
nuclear scales that are involved, the expression for the matrix element,
$\Scw$,
simplifies, in our model, to
\label\wexpression
\eq \Scw \simeq - \; { w_0 \pf \EF \over (2 \pi)^3} \;\; {\mee \over w_e}. \eeq
Using this expression, we tabulate
 in Table I the values that are required for $\mee/w_e$
to reproduce the observed anomalous signal. Following Ref.~\jim, we
integrate the anomalous rate only above an energy, $E_{\rm th}$, which
represents the point above which the contribution of the conventional
two-neutrino decay, $\bbtn$, is negligible.

\midinsert
$$\vbox{\tabskip=0pt \offinterlineskip
\halign to \hsize{\strut#& #\tabskip 1em plus 2em minus .5em&
\hfil#\hfil &#& \hfil#\hfil &#& \hfil#\hfil &#&
\hfil#\hfil &#& \hfil#\hfil &#\tabskip=0pt\cr
\noalign{\hrule}\noalign{\smallskip}\noalign{\hrule}\noalign{\medskip}
&& Element && $Q$ (MeV) && $E_{\rm th}$ (MeV) && $T^a_{1/2}$(yr) &&
$\mee/w_e$ &\cr
\noalign{\medskip}\noalign{\hrule}\noalign{\medskip}
&&\Ge\ && 2.0396 && 1.5 && \SNt{5}{22} && \SNt{8}{-5} &\cr
&&\Se\ && 2.995 && 2.2 && \SNt{5}{21} && \SNt{7}{-5} &\cr
&&\Mo\ && 3.0328 && 1.9 && \SNt{3}{20} && \SNt{2}{-4} &\cr
&&\Nd\ && 3.3671 && 2.2 && \SNt{3}{19} && \SNt{5}{-4} &\cr
\noalign{\medskip}\noalign{\hrule}\noalign{\smallskip}\noalign{\hrule}
}}$$

\centerline{\bf Table I} \smallskip
\noindent {\eightrm The Effective Yukawa coupling strength that is
required to reproduce the anomalous events in double beta decay. $\ss
T^a_{1/2}$
is the half-life of the anomalous events only, defined to be the total
number of events above a threshhold value, $\ss E_{\rm th}$, for the sum of
electron energies, above which essentially only excess events appear. $\ss
\mee/w_e$
is the coupling (defined in eq.~\lowmass) that is required to explain
the excess rate.}
\endinsert

It is clear from the table that the required values for $\mee/w_e$ are quite
consistent with the requirements --- $\mee \simeq 0.1 \,\eV$ and $w_e \simeq 1
\,\keV$ --- that were found earlier as being required to solve the solar- and
atmospheric-neutrino deficits.

Notice also that all four scalar states, and not just the two \ngb s,
contribute
to this decay since the energies available, $Q \simeq \MeV$, are much larger
than
the typical scalar masses, $w_i \simeq 1 \, \keV$. It is also important for the
above analysis that the scale, $\Lambda$, of the new physics which is
responsible for the effective interactions we are using, \cf\ eq.~\effdimsix,
is heavier than the nuclear physics scale, $\pf \simeq 60 \, \MeV$, of the
virtual neutrinos in the scalar-emitting $\bb$ decay. This is not just a
technical
requirement because, as we shall see, once a renormalizable model is
constructed, the $\bb$ decay rate becomes suppressed if all neutrino
states should be light compared to the MeV scale  \jim. As a result, an
explanation of the electron excess within the present framework ultimately
requires some new degrees of freedom that are not light on the scales that
are relevant for double beta decay.

\section{ A Renormalizable Model }

While the heavy mass scale, $\Lambda$, in our effective theory is large
compared
to the scale of neutrino physics, it is not necessarily large compared to the
weak scale. In order to verify that our model does not contradict standard
electroweak phenomenology and standard cosmology (in particular
nucleosynthesis)
we need to construct a renormalizable model for the interactions between the
new particles and SM particles at higher energies.  This is the
purpose of the present section. It is also only once this underlying physics
has been modelled that we may investigate the naturalness of this scheme. We
emphasize, however, that it is the effective theory presented in
 section 2 that explains the anomalous experimental results and any
renormalizable model that leads to it at low energies will do.

Consider, in addition to the scalar fields discussed to this point,
four left-handed electroweak-singlet Weyl fields, $s^\pm_e$ and $s^\pm_\mu$,
which respectively carry the charges $(\pm 1,0)$ and $(0,\pm 1)$ under \uiui.
The most general renormalizable couplings between these and the usual
standard-model particles are:
\label\renorm
\eq \eqalign{
&g_e s^-_e \psi_e H + g_\mu s^-_\mu \psi_\mu H
+ {h_e \over 2} s^+_e s^+_e \phi_e + h_\mu s^+_e s^+_\mu \phi_{e\mu}
+ {\ol{h}_e \over 2} s^-_e s^-_e \phi_e^* + \ol{h}_\mu s^-_e s^-_\mu
\phi^*_{e\mu} \cr
& \qquad \qquad + m_{s_e} s^+_e s^-_e + m_{s_\mu} s^+_\mu s^-_\mu
+ {\rm h.c.}, \cr}
\eeq
which reduces in unitary gauge to:
\label\apreshvev
\eq \eqalign{
&g_e s^-_e \nu_e (v+ \rho) + g_\mu s^-_\mu \nu_\mu (v+\rho )
+ {h_e \over 2} s^+_e s^+_e \phi_e +h_\mu s^+_e s^+_\mu \phi_{e\mu}
+ {\ol{h}_e \over 2} s^-_e s^-_e \phi_e^* + \ol{h}_\mu s^-_e s^-_\mu
\phi^*_{e\mu} \cr
& \qquad \qquad  + m_{s_e} s^+_e s^-_e + m_{s_\mu} s^+_\mu s^-_\mu +   {\rm
h.c.}, \cr} \eeq
if $\rho$ denotes the physical SM Higgs.

Since the lepton-breaking {\it v.e.v.}'s, $w_i$, are so small, we may as a
first
approximation neglect them. Then the neutrino mass matrix breaks up into two
three-by-three submatrices for both the electron- and muon-neutrino
sectors. These submatrices have the form:
\label\renormatrix
\eq \pmatrix{0 & m_{s_i} & 0 \cr
	m_{s_i} & 0       & g_i v \cr
	0       &g_i v    & 0 \cr},
\eeq
where $i=e,\mu$. Each has one vanishing eigenvalue, and a pair of nonzero
ones: $\pm M_{s_i} \equiv \pm\sqrt{m_{s_i}^2 + (g_i v)^2}$. The corresponding
eigenvectors are
\label\eigenv
\eq
S_i^\pm = {1 \over \sqrt2 M_{s_i}} \pmatrix{m_{s_i} \cr \pm M_{s_i} \cr
 g_i v \cr} \qquad \hbox{and} \qquad
\nu_{i0} = {1 \over M_{s_i}} \pmatrix{ -g_iv \cr 0\cr
m_{s_i} \cr}. \eeq
This gives the following relation between the weak-interaction and mass
eigenstates:
\label\weaktomass
\eq
\pmatrix{s_i^+ \cr s_i^- \cr \nu_{i} \cr}  = {1 \over \sqrt2 M_{s_i}}
\left[ \matrix{ m_{s_i} & m_{s_i} & - \sqrt2 g_i v \cr
M_{s_i} & - M_{s_i} & 0 \cr g_iv & g_iv & \sqrt2 m_{s_i} \cr} \right]
\pmatrix{ S_i^+ \cr S_i^- \cr \nu_{i0} \cr}.
\eeq

If the heavy Dirac fields, $S^\pm_i$, are integrated out, then the effective
interaction of eq.~\effdimsix\ is obtained with:
\label\correspondence
\eq
{\lambda_e \over \Lambda^2} = {g_e^2 h_e \over M_{s_e}^2}, \qquad \hbox{and}
\qquad {\lambda_\mu \over \Lambda^2} = {g_e g_\mu h_\mu \over M_{s_e} M_{s_\mu}
}. \eeq
Notice that the more weakly coupled these sterile neutrinos are, the lighter
they must be to preserve the desired size of $\mee$ and $\memu$, (recall
$\mee/w_e=\lambda_e v^2 / \Lambda^2$ and
$\memu/w_\mu=\lambda_\mu v^2 / \Lambda^2$).

It is worth recalling, in this regard, that if the electron-type sterile
neutrino, $S_e^\pm$, is much lighter than several MeV, then the predicted
anomalous $\bb$ decay rate becomes suppressed, since in this limit the result
is proportional to the $\nu_e - \nu_e$ element of the full renormalizable mass
matrix, which is zero. This is a special case of the more general suppression,
discussed in Ref.~\jim, of the anomalous double beta decay rate in the
low-mass limit. We are therefore led to prefer $M_{s_e} \gsim 60 \, \MeV$.

A useful quantity for comparison with phenomenological constraints is the
mixing angle which controls the strength of the participation of the sterile
states in the charged-current weak interactions. From eq.~\weaktomass\ we see
that the sterile mixing into the electron and muon charged currents are given
by:
\label\mixing
\eq
V_{eS_e^+} = V_{eS_e^-} = {g_e v \over \sqrt2 M_{s_e}}, \qquad \hbox{and}
\qquad
V_{\mu S_\mu^+} = V_{\mu S_\mu^-} = {g_\mu v \over \sqrt2 M_{s_\mu}}. \eeq
These combinations are useful because they are typically constrained to be
small, with the strength of the bound depending on the mass of the heavy
neutrino state, and they are bounded from below in the models we
consider. Taken together these exclude some of the potential mass ranges for
the
sterile neutrinos.

In the MeV mass range the experimental constraints come from $\pi$ and $K$
meson
decays. For example, measurements of $\Gamma(\pi\to e\nu)/\Gamma(\pi\to
\mu\nu)$
strongly bound the mixing into the electron charged current for neutrinos in
the
particularly interesting mass range between 1 and 100 MeV. Such a mixing with a
1~MeV neutrino must satisfy $|V_{eS}|^2 < 10^{-3}$ at the 90\% C.L., whereas
the
same bound for a 10~MeV and a 50~MeV neutrino is $10^{-5}$ and \SNt{5}{-7},
respectively.
\ref\brit{D.I.Britton \etal, \prd{46}{92}{R885}.}
%
\brit. Searches for a nonstandard component to $K \to e \nu$
\ref\kaonbound{T. Yamazaki \etal, in Proc. XIth Intern. Conf. on Neutrino
Physics and Astrophysics, K. Kleinknecht and E.A. Paschos (World Scientific,
Singapore, 1984) p. 183.}
\kaonbound\ extend this limit up to neutrino masses of 350 MeV.

Somewhat
weaker constraints on sterile neutrino mixing with $\nu_\mu$ come from
measurements of the muon spectrum in $\pi \to \mu\nu$
\ref\gilman{A summary of  bounds on heavy neutrino mixings
may be found in, F.J. Gilman,
 {\it Comments Nucl. Part. Phys.} {\bf 16} (1986) 231.}
\gilman.
For example a 1~MeV neutrino must satisfy
$|V_{\mu S}|^2 < 10^{-2}$ at the 90\% C.L., and mixing with
a 10~MeV neutrino must be less than $10^{-5}$
\ref\sinmu{ David I. Britton private communication}
\sinmu.
It is noteworthy that beam-dump bounds do not apply to our model,
since these experiments typically search for neutrino decays into charged
particles, and our neutrino eigenstates predominantly decay invisibly into
neutrinos and light scalars.

In our model, a lower bound on these parameters starts from the following
relation~(see eqs.~\correspondence,\mixing):
\label\lowerbound
\eq |V_{eS_e}|^2 \equiv | V_{eS_e^+}|^2 + | V_{eS_e^-}|^2 = {\mee \over w_e
h_e}.
\eeq
This, when taken together with the value $\mee/w_e \approx 10^{-4}$, and the
perturbative limit, $h_e < 4 \pi$ , leads to the inequality $|V_{eS_e}|^2 \gsim
8 \times 10^{-6}$. Thus $M_{s_e}\lsim$ 10~MeV.
Since these searches extend to masses of order 350~MeV,
we've excluded $M_{s_e}$ in the range 10--350~MeV.
Likewise $|V_{eS_e}||V_{\mu S_\mu}|=\memu/(w_\mu h_\mu) \gsim
8 \times 10^{-6}$.

In summary these bounds tell us is that the phenomenologically acceptible
parameter range is the one for which $g_i v \ll M_{s_i} \simeq m_{s_i}$.
With this model in hand, we may now confront the remaining phenomenological
constraints, as well as the naturalness issue.

\section{ Cosmological Bounds}

The standard Big-Bang model of cosmology and primordial nucleosynthesis
furnishes strong constraints on any model which contains light neutrinos
and scalars, as do our models
\ref\Chang{S.~Chang and K.~Choi,  Seoul preprint SNUTP 92-87.}
\Chang. The success of Big-Bang
nucleosynthesis sets an upper bound to the number of gravitating degrees of
freedom at the time when the photon temperature is of order $T_\gamma \sim (0.1
-
2) \, \MeV$. The more degrees of freedom there are at this point, the faster
the
universe expands, and so the more neutrons are available to be cooked into
${}^4$He. This raises the predicted primordial mass fraction, $Y_P$, of
${}^4$He.
Since present limits require $ 0.22 \le Y_P \le 0.24$ at the 95\% C.L.
\ref\WSSOK{ T.P.~Walker, G.~Steigman, D.N.~Schramm, K.A.~Olive and H.S.~Kang,
Astrophys. J. {\bf 376} (1991) 51.}
\WSSOK, and since this agrees with what is expected for the standard model
particle content, there is little room for new particles to be abundant at this
temperature. It is conventional to express the resulting
bound as a limit on the number of neutrinos,
 $1.3 < N_{\rm eff} = N_\nu + \delta N < 3.2$, where the
standard-model value is $N_\nu = 3$, and the observationally preferred value is
$N_\nu=2.2$.

Every new spin-half or spinless relativistic particle that is present when the
photon temperature is $T_\gamma$ contributes to $\delta N$ an amount:
\label\numnus
\eq  \delta N = \sum_{\rm fermions} (T_i/T_\gamma)^4 +
{4 \over 7}  \sum_{\rm bosons} (T_i/T_\gamma)^4 , \eeq
where $T_i$ is the temperature of particle type `$i$'. Since $T_i = T_\gamma$
for particles that are in equilibrium with photons, any such particles in
addition to the usual three neutrinos are basically ruled out. For two light
complex scalars, for example, $\delta N = 16/7$, which is clearly too large.

There are two standard ways to avoid this bound. ($i$) The above expression
does not apply to particles which are nonrelativistic and still in equilibrium,
since for temperatures below their rest masses the abundance of such particles
is suppressed by the Boltzmann factor, $e^{-m/T}$. ($ii$) Alternatively, if a
particle decouples from the photon bath at sufficiently early times -- in
practice this means before the QCD transition at
$T_{\sss QCD}\sim (0.1 - 1)$~GeV
-- they can be diluted by subsequent reheating of the photons. In this case the
ratio $T_i/T_\gamma$ can become sufficiently small to suppress this particle's
contribution to $\delta N$.

Only the first of these mechanisms is available to us here, and this
only for the potentially heavy particles such as $S_i^\pm$.
Scenario ($ii$) can not occur in our model since all of
the heavy and light neutrinos are in equilibrium with one another due to the
exchange of the light scalars, $\phi_i$. (In fact for
temperatures above $T_\nu \simeq 2.3$ MeV,
all of these particles are also in equilibrium with ordinary matter
because of the weak interactions of the ordinary neutrinos.)

As a result, any
heavy particle in the neutrino sector whose mass satisfies $m > T_\nu$
satisfies the conditions of loophole ($i$) above. As $T$ falls below its mass,
it annihilates out, reheating both the neutrino and ordinary sectors, and
therefore not contributing to $\delta N$. Any of our new particles that are
heavier than a few MeV are therefore cosmologically benign, due to their
equilibrating interactions involving the light scalars.

Neither option can apply to the light scalars themselves, however, and these
particles already contribute unacceptably to $\delta N$. Another mechanism is
required in order to suppress these. One possibility arises if one of the
heavier states should have a mass, $M$, that lies below
$T_\nu \simeq 2.3$ MeV, but above the
temperature, $T_{n/p} \simeq 0.7$ MeV, at which the neutron-proton ratio
freezes out. In this case, when $T$ falls below $M$, annihilation into
light scalars raises the temperature of the neutrino sector, thereby
increasing, in particular, the number of $\nu_e$'s. As is pointed out in
\ref\negneutrino{K. Enquist, K. Kainulainen and M. Thomson, \prl{68}{92}{744}.}
Ref.~\negneutrino, however, an enhanced $\nu_e$ density at this time
suppresses the neutron abundance, and so decreases the amount of ${}^4$He
that is ultimately produced. Quantitatively, if $n/n_0$ and $E/E_0$
respectively
denote the ratio of the $\nu_e$ number densities and energies before and after
annihilation, then the excess $\nu_e$'s (and $\ol{\nu}_e$'s) suppress $Y_P$ by
an amount equivalent to $\delta N = - 4.6 \delta n$, where $\delta n = (n/n_0)
(E/E_0)^2 -1$.

We next attempt to estimate the implications of this effect for our models.
Suppose, therefore, that $N_F$ species of Weyl fermions and $N_S$ real scalars
should annihilate when $T_{n/p} < T < T_\nu$, leaving $n_F$ and $n_S$ fermions
and scalars at low temperatures. Then the increase in $\nu_e$ energy and
density
due to the annihilation of these particles may be estimated to be $(n/n_0) =
(T_{\rm after} / T_{\rm before})^3$ and $(E/E_0) = (T_{\rm after} /
T_{\rm before})$, where entropy conservation requires  $(T_{\rm after}/ T_{\rm
before})^3 = [( n_F + N_F) + {4\over 7} \; (n_S + N_S) ]/[ n_F + {4\over 7}
\; n_S]$. As a result we find
\eq
\delta n = \left[ {( n_F + N_F) + {4\over 7} \; (n_S + N_S) \over n_F + {4\over
7}  \; n_S} \right]^{5/3} - 1. \eeq

For example, for the models under consideration there are three light neutrinos
and two complex scalars at low energies, so $n_F = 3$ and $n_S = 4$. If one of
our massive Dirac neutrinos, $S_i^\pm$, should annihilate out at this strategic
time then $N_F = 2$ and $N_S =0$. Then this leads to $\delta n = 0.71$
and $N_{\rm eff} = 2.0$. This is well within the experimentally permitted
range, which actually prefers fewer than three neutrinos!

For future reference we also comment on the consistency of the
supersymmetric model with cosmological constraints.  In such a model
 the two light scalars and 1.5 MeV Dirac neutrino as well as their
(approximately degenerate) superpartners must be added to the standard three
neutrinos. In this case we have $n_F = 5$, $n_S = 4$, $N_F = 2$ and $N_S
= 4$. We thus find $\delta n = 1.2$ and $N_{\rm eff} = 1.9$.
 There is also the gravitino in the supergravity version of the model.
It has mass of order 1 keV or smaller and thus satisfies the
cosmology bound necessary to prevent it from closing the universe.
This bound states that $m_{\rm gravitino}< 1$ keV or $>$10 TeV,
\ref\pagelsprimack{H. Pagels and J. Primack, \prl{48}{82}{223}.}
\pagelsprimack,
\ref\weinbergravitino{S. Weinberg, \prl{48}{82}{1303}.}
\weinbergravitino.

In summary, we see that if this is the mechanism for evading the
nucleosynthesis bound, then cosmology argues for having one of the
sterile Dirac neutrinos with a mass of around 1.5 MeV, with the second
sterile Dirac state being heavier.  This heavier state should be the
electron-type state, $S_e^\pm$, if the double beta decay anomaly is to
be reproduced.  $S_e^\pm$ must be very heavy in order to avoid the
$\pi$- and $K$-decay bounds.  In both the supersymmetric and
nonsupersymmetric cases we are led to the prediction $N_{\rm eff}
\simeq 2$ for primordial nucleosynthesis.  Finally the gravitino
 in the supersymmetric model has mass $\roughly{<}$ 1 keV and hence
satifies cosmological constraints.

\section{ Naturalness }

We have seen that both the oscillation--decay solution for both the solar- and
atmospheric neutrino problems, and the scalar-emission explanation of
the excess double beta decay events, point to the necessity for scalar {\it
v.e.v.}'s of about 1 keV. Generically a small scalar mass and expectation value
are unacceptable because they are not stable under renormalization down from
higher scales. We wish, in this section, to quantify this statement and to
determine what constraints are required for the parameters of the theory in
order to acheive this stability, or naturalness.

The small parameters which are required to ensure such small scalar masses and
{\it v.e.v.}'s appear in the scalar potential, which in the present model has
the following renormalizable form:
\label\potential
\eq \eqalign{
V(\phi_e,\phi_{e\mu},H) &= -\mu_e^2 | \phi_e |^2 - \mu_\mu^2 | \phi_{e\mu} |^2
- \mu_{\sss H}^2 H^\dagger H + \xi_{ee} |\phi_e|^4 + \xi_{\mu\mu}
|\phi_{e\mu}|^4 + \xi_{e\mu} |\phi_e|^2 |\phi_{e\mu}|^2 \cr
& \qquad \qquad + \xi_{e{\sss H}} |\phi_e|^2 H^\dagger H + \xi_{\mu{\sss H}}
|\phi_{e\mu}|^2 H^\dagger H + \xi_{\sss HH} (H^\dagger H)^2. \cr} \eeq
The hierarchy problem arises because, whereas $\mu_{\sss H} \sim 100$~GeV,
$\mu_e$ and $\mu_\mu$ can only be $\sim 1 \, \keV$. Similarly, although this
hierarchy does not preclude some of the dimensionless couplings, namely
$\xi_{\sss HH}$, $\xi_{ee}$, $\xi_{e\mu}$ and $\xi_{\mu\mu}$ from being $O(1)$,
those which couple the light to the heavy fields, $\xi_{e{\sss H}}$ and
$\xi_{\mu{\sss H}}$, cannot be larger than $\xi_{i{\sss H}} \lsim
O(\mu_i^2/\mu_{\sss H}^2) \sim 10^{-16}$. Otherwise quartic interactions such
as
$|\phi_e|^2 H^\dagger H$ would generate large mass terms for $\phi_e$ once $H$
receives its {\it v.e.v.} of 174 GeV.

We must ask whether such small values for the couplings are ruined when they
are run down from higher scales. For definiteness we define our running within
the `decoupling-subtraction' renormalization scheme, using dimensional
regularization
\ref\decoupling{For a very clear description of this scheme, see
I. Hinchliffe, Collider physics for the late 1980's,
in {\it TASI 86 Lectures in Elementary Particle Physics},
Inst.\ Santa Cruz, CA June 23 -- July 19, 1986.
}
\decoupling. In this scheme all couplings are run between particle threshholds
using modified minimal subtraction, and each particle is integrated out as the
renormalization point is lowered below its threshhold. Within this framework,
the \msbar\ running between threshholds only introduces a
logarithmic dependence on heavy mass scales, and the potentially dangerous
powers of heavy scales arise when these heavy particles are integrated out.

There are several kinds of graphs to consider, depending on which particles
circulate in the loop. The most dangerous ones are those which involve the
heaviest particles that are coupled the most strongly to the light scalars.
\topic{Higgs Loops} It is fairly easy to see that the running of the small
couplings due to loops involving the standard-model Higgs are just as small as
are the couplings themselves. For example the graphs of Figs. (1) give the
following contributions to $\delta \mu_i^2$ and $\delta \xi_{i{\sss H}}$:
\label\higgschange
\eq
(\delta\mu_i^2)_{\rm Fig.~1} \sim {\xi_{i{\sss H}} M_{\sss H}^2 \over 16 \pi^2}
\sim (0.1 \, \keV)^2, \qquad \hbox{and} \qquad
(\delta \xi_{i{\sss H}} )_{\rm Fig.~1} \sim {\xi_{i{\sss H}} \xi_{\sss HH}
\over
16 \pi^2} \sim 10^{-18}, \eeq
where we take $M_{\sss H} \sim 100$ GeV, and we ignore all logarithms of large
mass ratios.
\topic{Neutrino Loops}
The really dangerous graphs involve heavy neutrinos, since the couplings of
these particles to the light scalars is constrained by the requirement that a
sufficiently large $\nu_e - \nu_\mu$ mass matrix be generated at low energies.
For example, the graphs of Fig.~(2) produce the following contribution to
 $\delta \mu_e^2$ and $\delta \xi_{e{\sss H}}$:
\label\neutrinochange
\eq
(\delta\mu_e^2)_{\rm Fig.~2} \sim {h_e^2 M_{s_e}^2 \over 16 \pi^2}, \qquad
\hbox{and} \qquad  (\delta \xi_{i{\sss H}} )_{\rm Fig.~2} \sim { h_e^2 g_e^2
\over 16 \pi^2}, \eeq
with similar contributions to $\delta \mu_\mu^2$ and $\delta \xi_{\mu{\sss
H}}$.
\endtopic

\midinsert
$$\vbox{\tabskip=0pt \offinterlineskip
\halign to \hsize{\strut#& #\tabskip 1em plus 2em minus .5em&
\hfil#\hfil &#& \hfil#\hfil &#\tabskip=0pt\cr
\noalign{\hrule}\noalign{\smallskip}\noalign{\hrule}\noalign{\medskip}
&& Source && Constraint &\cr
\noalign{\medskip}\noalign{\hrule}\noalign{\medskip}
&& Weak Universality && $g_e \lsim \SN{4.5}{-4} \; m_{350}$ &\cr
&& && \cr
&& $\pi \to \mu \nu$ && $g_\mu \lsim \SN{8.6}{-7} \; m_{1.5}$ &\cr
&& && \cr
&& Naturalness ($\delta \mu^2_i$) && $h_e, h_\mu \lsim \SN{4}{-5} \;
m_{350}^{-1}$ &\cr
&& && &\cr
&& Naturalness ($\delta \xi_{i{\sss H}}$) && $g_e h_e, g_e h_\mu, g_\mu h_\mu
\lsim \SN{1}{-7}$ &\cr
&& && \cr
&& $\nu_e - \nu_\mu$ Mass Matrix && $g_e^2 h_e \simeq \SN{4}{-10} \;
m_{350}^2$ &\cr
&&  && $g_e g_\mu h_\mu \simeq \SN{2}{-12} \; m_{350} \, m_{1.5}$ &\cr
\noalign{\medskip}\noalign{\hrule}\noalign{\smallskip}\noalign{\hrule}
}}$$

\centerline{\bf Table II} \smallskip
\noindent {\eightrm A summary of the constraints on the
scalar-neutrino couplings. $\ss m_{350}$ denotes $\ss (M_{s_e}/ 350 \, \MeV)$
and
$\ss m_{1.5}$ represents $\ss (M_{s_\mu}/ 1.5 \, \MeV)$ .}
\endinsert

We see here the difficulty with having neutrinos in the MeV mass range. We
work with the neutrino masses of most interest for double-beta decay and
nucleosynthesis, $M_{s_e} \simeq 350$ MeV and $M_{s_\mu} \simeq 1.5$ MeV.
 We've taken $M_{s_e}\roughly{>}350$  to avoid all bounds from $K\to
e\nu$ and $\pi \to e \nu$ decays \kaonbound.
The mixing of such a massive neutrino with
$\nu_e$ is then constrained principally from weak universality, and satsifies
$|V_{eS_e}|^2 \lsim 0.05$ \gilman.
 The bounds we obtain are listed in Table II. From
this table we see that even if all couplings are made as large as they can be
while remaining consistent with naturalness (and direct phenomenological
constraints), their contributions to
$\nu_e - \nu_\mu$ mass matrix
still fall two orders of magnitude short of what is required.

The naturalness constraint can be eased by allowing the neutrinos to become
lighter, since this allows their couplings to become weaker while keeping the
desired low-energy $\nu_e - \nu_\mu$ mass matrix fixed. This can only be done
at the expense of reintroducing the cosmological problems and giving up
our explanation for the double-beta
decay anomaly.  We consider a different approach in the next section.

\section{ A Supersymmetric Model }

An alternative approach is to hold the line on the heavy-neutrino masses, and
to instead try to work within a framework for which the naturalness conditions
take a weaker form. There are two ways known to make a very light scalar {\it
v.e.v.} natural. One either demands that the light scalar be a composite
of underlying nonscalar particles, or one arranges for cancellations between
fermions and bosons in the renormalization of small parameters. These
cancellations can be ensured naturally if the model is supersymmetric.

We have tried both approaches, but have only found a workable example in the
supersymmetric class. Here the contributions of the superpartners of each of
the particles introduced so far largely cancel in their contributions to
$\delta \xi_{i{\sss H}}$ and $\delta \mu_i^2$. This cancellation is only
accurate enough for scalar particles at the keV scale if the relevant
suprmultiplets are themselves split in masses only by roughly a keV. Yet this
must be made consistent with the fact that the superpartners of the charged
leptons and quarks must have masses that are of order the electroweak scale.

This may be consistently done within a supersymmetric framework if
these highly degenerate multiplets are much more weakly coupled to the
supersymmetry-breaking sector than are the other garden-variety
particles we know. We demonstrate that this is possible in this
section by presenting an existence proof, in the form of a
supersymmetric extension of our renormalizable model of Section (4).
We shall find that, within this extension, the naturalness
requirements are significantly relaxed.

\subsection{The Model}

The model we shall consider consists of the straightforward supersymmetric
extension of our renormalizable model of Section (4). We imagine,
therefore, supplementing the supersymmetric standard model (SSM) with four
electroweak-singlet left-chiral superfields, carrying the following \uiui\
charge assignments:
\label\susyrep
\eq  \eqalign{
\Phi_e \ni \{ \phi_e , \chi_e \} \sim (-2,0), \qquad&\qquad
\Phi_{e\mu} \ni \{ \phi_{e\mu} , \chi_{e\mu} \} \sim (-1,-1), \cr
N_e^\pm \ni \{ \Scn_e^\pm , s^\pm_e \} \sim  (\pm 1, 0) , \qquad&\qquad
N_\mu^\pm \ni \{ \Scn_\mu^\pm , s^\pm_\mu \} \sim (0,\pm1). \cr} \eeq
Our notation here lists the scalar ($\phi_e$) and spinor ($\chi_e$) components
of each superfield ($\Phi_e$). The most general renormalizable \uiui-invariant
superpotential becomes:
\label\rensuperpot
\eq \eqalign{
W &= W_{\sss SSM} + g_e \Psi_e N^-_e H_u + g_\mu \Psi_\mu N^-_\mu H_u +
{ h_e \over 2} N^+_e N^+_e \Phi_e +  h_\mu N^+_e N^+_\mu \Phi_{e\mu} \cr
& \qquad \qquad + m_{s_e} N_e^+ N_e^- + m_{s_\mu} N_\mu^+ N_\mu^-, \cr} \eeq
where $\Psi_i \ni \{ \Scs_i, \psi_i \}$ represent the SSM left-handed
lepton-doublet superfields, which transform under \uiui\ as $\Psi_e \sim (1,0)$
and $\Psi_\mu \sim (0,1)$. $H_u$ similarly represents the SSM Higgs superfield
whose {\it v.e.v.} gives up-type quarks their masses.

This model as it stands does not break supersymmetry, and in the limit of
negligible lepton-number breaking {\it v.e.v.}'s for the scalar fields,
$\phi_e$,
$\phi_{e\mu}$, $\Scn_e^\pm$ and $\Scn_\mu^\pm$, it predicts completely
degenerate supermultiplets having the following masses: Four massless
multiplets ($\Phi_e$, $\Phi_{e\mu}$, and one linear combination of $N^+_i$ and
the neutrino multiplet in $\Psi_i$, for both $i=e$ and $\mu$); a pair of
degenerate massive multiplets ($N^-_e$ and the other combination of
$N^+_e$ and the neutrino multiplet in $\Psi_e$) having masses $M_{s_e} =
\sqrt{m_{s_e}^2 + g_e^2 v^2}$; and a similar pair for the muon-neutrino sector.
Notice that, because of the unbroken supersymmetry, the scalars $\phi_e$ and
$\phi_{e\mu}$ are massless, regardless of how large the couplings $g_e$ and
$h_e$ should be.

All of the interactions that are required to account for the
solar- and atmospheric-neutrino problems, as well as the double-beta decay
excess, appear in the superpotential of eq.~\rensuperpot\ once this model is
written out in terms of its component fields. This is because this
superpotential
contains a counterpart for each of the terms in the renormalizable couplings of
eq.~\renorm, {\it except} for those terms which involve $\phi_e^*$ or
$\phi_{e\mu}^*$ coupled to left-handed fields. Since these last terms do not
appreciably affect the $\nu_e - \nu_\mu$ sector, this difference is not
important for our purposes.

\subsection{Supersymmetry Breaking}

In order to proceed further, some properties must be assumed for the
supersymmetry-breaking sector of the model. Most importantly, it is necessary
to determine whether it is possible to break supersymmetry in such a way as to
give electroweak-scale masses to the superpartners of the ordinary quarks,
leptons and gauge fields, and yet to still keep the mass-splittings within the
supermultiplets $\Phi_e$, $\Phi_{e\mu}$, $N^\pm_e$ and $N^\pm_\mu$ sufficiently
small to permit naturally small expectation values $\avg{\phi_e} \sim
\avg{\phi_{e\mu}} \sim 1~\keV$.

In order to address this question, we imagine that supersymmetry is
spontaneously broken when the auxiliary field of some electroweak- and \uiui\
singlet superfield, $Y$, develops an expectation value: $\avg{Y} = \Lambda_s^2
\theta_\lft^2$, where $\theta_\lft$ is the Grassman coordinate of $Y$. Since
$\avg{Y}^2 =0$, the most general nonderivative couplings this expectation value
may have with the other fields are given by replacing $Y \to \avg{Y}$ in the
following K\"ahler and superpotentials:
\label\susybreaking
\eq
\delta K = A(X,X^*) \; Y^* Y + [ B(X,X^*) \; Y^* + \cc ], \qquad \hbox{and}
\qquad  \delta W = C(X) \; Y. \eeq
In these expressions $X^a \ni \{\Scx^a, \chi^a \}$ collectively represent all
of the superfields of the SSM as well as the additional supermultiplets that
were
introduced in eq.~\susyrep\ above. Once $Y$ is replaced by $\avg{Y}$, $\delta
K$
and $\delta W$ reduce to a set of soft SUSY-breaking contributions to the
scalar
potential, $\delta V$, as well as to the Yukawa-couplings, $\delta \Scl_{\rm
yuk}$. If $F_x^a$ represents the auxiliary field for supermulitplet $X^a$, then
these contributions may be written
\label\components
\eq \eqalign{
\delta V(\Scx,\Scx^*) &= \Lambda_s^4 \; A(\Scx,\Scx^*) + \Lambda_s^2 \left[
{\partial B \over \Scx^a} \, F_x^a + C(\Scx) + \cc \right] , \cr
\delta \Scl_{\rm yuk} &= {\Lambda_s^2 \over 2} \; \left( {\partial^2 B \over
\partial \Scx^a \partial \Scx^b}\right) \; \chi^a \Pl \chi^b + \hbox{(gaugino
terms)} + \cc. \cr} \eeq

We adopt the usual `hidden-sector'
\ref\hiddensector{J. Polchinski and L. Susskind, \prd{26}{82}{3661};
For a review see H.P. Nilles, \prep{110}{84}{1}.}
scenario \hiddensector\ in which the SUSY-breaking field, $Y$, couples to
electroweak-scale fields only through interactions which are suppressed by a
large mass scale, $M$. We may then write the most general lowest-dimension
contributions which involve
the non-SSM fields as:\foot\sugra{If {$\ss M$} should be
as large as the Planck scale, then there would also be additional terms which
arise from the elimination of the auxiliary fields of the supergravity
multiplet.}
\label\lowestdim
\eq\eqalign{
\delta A &= {1 \over M^2 } \left[ a_e |\Phi_e|^2 + a_\mu
|\Phi_{e\mu}|^2 + \left( a'_e N_e^+ N_e^- + a'_\mu N_\mu^+ N_\mu^-
+ \cc \right) + \sum_{i=e,\mu} \sum_{n=\pm} a_i^n |N_i^n|^2 \right],\cr
\delta B &= {1 \over M } \left[ b_e |\Phi_e|^2 + b_\mu
|\Phi_{e\mu}|^2 + \left( b'_e N_e^+ N_e^- + b'_\mu N_\mu^+ N_\mu^-
+ \cc \right) + \sum_{i=e,\mu} \sum_{n=\pm} b_i^n |N_i^n|^2 \right], \cr
\delta C &= c_e N_e^+ N_e^- + c_\mu N_\mu^+ N_\mu^-  \cr
&\qquad \qquad + {1 \over M } \left[ \tw{g}_e \Psi_e N_e^- H_u + \tw{g}_\mu
\Psi_\mu N_\mu^- H_u +  {\tw{h}_e \over 2} N_e^+ N_e^+ \Phi_e + {\tw{h}_\mu
\over 2} N_\mu^+ N_\mu^+ \Phi_{e\mu} \right]. \cr}
\eeq
Using these expressions in eqs.~\components\ gives the induced
SUSY-breaking interactions for our model. The induced scalar mass terms which
arise in this way are generically $O(\Lambda_s^2/M)$, with the exception of the
terms $c_e N_e^+ N_e^- + c_\mu N_\mu^+ N_\mu^- $ in $\delta C$, which induce
masses for $\Scn_i^\pm$ that are $O(\Lambda_s)$. Since this latter size is much
too large --- as is the corresponding SSM term $\propto H_u H_d$ --- we take
$c_e = c_\mu = 0$. Since $\delta C$ is a contribution to the superpotential,
this choice is technically natural due to the celebrated perturbative
nonrenormalization theorems of supersymmetry.

Generally the heavy mass scale, $M$, is chosen to ensure that the generic
scalar masses that are obtained for the superpartners of the quarks and leptons
are of order the weak scale, $\Lambda_s^2/M \sim v$. With this choice the
induced masses for scalars like $\phi_e$ would also be $O(v)$, which is
unacceptably large. We must therefore suppose that all contributions to $\delta
A$ through $\delta C$ are much smaller than would be indicated simply by the
powers of $M$ in eqs.~\lowestdim. We choose to parameterize this extra
suppression by taking $M$ to be much larger than the heavy mass,
$M_{\sss SSM}$,
which appears in the corresponding SSM terms (which we call
$A_{\sss SSM}$,
$B_{\sss SSM}$ and $C_{\sss SSM}$). This might be arranged in an underlying
theory by having all of the couplings between the supersymmetry-breaking sector
and the very light fields, $\Phi_i$ and $N_i^\pm$, involve the virtual exchange
of particles of mass $M$, while the coupling between $Y$ and the SSM fields
arises through the exchange of lighter particles of mass $M_{\sss SSM}$. $M$,
$M_{\sss SSM}$ and $\Lambda_s$ must therefore be related by:
$\Lambda_s^2/M
\simeq 1~\keV$ and $\Lambda_s^2/M_{\sss SSM} \simeq
100~\GeV$, so $M/M_{\sss
SSM} \simeq (100~\GeV/1~\keV) \simeq 10^8$.
If we choose, for example, $M \sim
M_{\sss P} \simeq 10^{19}~\GeV$, then $M_{\sss SSM}
\simeq 10^{11}~\GeV$ and
$\Lambda_s \simeq \SN{3}{6}~\GeV$.

The million-dollar question is whether this hierarchy of SUSY-breaking
terms is technically natural. The potentially difficult hierarchy to
maintain is that between the sizes of $\delta A$ and $A_{\sss SSM}$,
and of $\delta B$ and $B_{\sss SSM}$ since these are part of the
K\"ahler potential, and so are not protected by the nonrenormalization
theorems. This hierarchy would certainly be stable if all of the
couplings between SSM particles and the very light sector were
suppressed by factors like $M_{\sss SSM}/M$, but this is {\it not}
true for our model due to the comparatively large couplings, $g_i$ and
$h_i$, in the supersymmetric superpotential of eq.~\rensuperpot. We
must therefore check whether these interactions can communicate the
large SSM SUSY-breaking scale to the light supermultiplets.

Some of the potentially dangerous graphs are given in Fig. (3). We write these
graphs using the supergraph formulation, in terms of which the supersymmetric
cancellations are the most explicit. Fig. (3a) relates the splitting in the
$N_i^\pm$ supermultiplets to those in $\Phi_e$ and $\Phi_{e\mu}$, and gives:
\label\splita
\eq
\delta \left({a_e \over M^2}\right) \sim {h_e^2 \over 16 \pi^2} \; \left(
{a_e^+
\over M^2} \right), \eeq
from which we infer: $\delta \mu_e^2  \sim (h_e^2 /16\pi^2) \; (a_e^+
\Lambda_s^4/M^2)$. This has the same form as for the nonsupersymmetric model,
except $M_{s_e}^2$ has been replaced by the splitting within the $N_e^+$
supermultiplet: $\Delta M^2 \sim a_e^+ \,\Lambda_s^4/M^2$. This is naturally
small enough for our purposes, since the direct contribution $\mu_e^2 \simeq
a_e
\Lambda_s^4/M^2$ is by assumption $O(1~\keV)$. But it underlines the fact that
the members of the $N_i^\pm$ supermultiplets cannot be split in mass by more
than a few orders of magnitude more than are the very light states in $\Phi_e$
or $\Phi_{e\mu}$.

Fig. (3b) illustrates the other potentially dangerous combination, in which
it is the mass splitting, $M_\Scs^2=\Lambda^4_s(a_{\sss SSM}
/M_{\sss SSM}^2)$, between the SSM sneutrinos and neutrinos
that sets the scale of the important loop momenta. This graph produces the
following estimate for $\delta a_e$:
\label\sneutrinoloop
\eq  \delta \left( {a_e \over M^2} \right) \sim {g_e^2
m_{s_e}^2 h_e^2 v^2 \over
16 \pi^2 M_\Scs^4} \; \left({a_{\sss SSM} \over M_{\sss SSM}^2}\right), \eeq
from which we obtain: $\delta\mu_e^2 \sim (h_e^2/16\pi^2)\, (g_e v/m_{s_e})^2
\, (m_{s_e}/M_\Scs)^4 \; M_\Scs^2$
This has a simple interpretation in terms of ordinary graphs
involving the component fields. It reflects the fact that the
contribution of the massive sneutrino is proportional to $(h_{\rm eff}^2 /
16\pi^2) \; M_\Scs^2$, where $h_{\rm eff} \sim h (g_e v/m_{s_e})
(m_{s_e}^2/M_\Scs^2)$ is the effective coupling between the heavy sneutrino and
the light scalar. The factor of $g_e v/m_{s_e}$ reflects the mixing angle
between the SSM sneutrino multiplet, $\Psi_e$, and the multiplets, $N_e^\pm$,
which couple to $\phi_e$. The remaining factor, $m_{s_e}^2/M_\Scs^2$, similarly
reflects the SUSY-breaking mixing angle between the heavy sneutrino mass
eigenstate and the lighter scalar flavour eigenstates.

We see, then, that this is indeed a stable hierarchy of scales, because the
loop-induced splitting in the light multiplets due to the weak-scale splitting
of the SSM multiplets is sufficiently suppressed by the small coupling between
these two sectors.

\section{ Conclusions }

We have constructed several models which incorporate the
phenomenological oscillation/decay solution of Acker \etal\ \acker\
for the solar and atmospheric neutrino problems. In so doing we have
shown how the neutrino-scalar couplings that are required to produce
the desired rate for neutrino decay, are also of the strength that is
required to account for the recently-observed excess double beta decay
events. This excess would then be interpreted as the emission of light
scalars in addition to the electrons, with a predicted sum electron
spectrum that falls into the `Ordinary-Majoron' class of Ref.~\jim. We
regard this potential integration of the double-beta decay anomaly
with the mechanism underlying the solar and atmospheric problems to be
of sufficient interest to justify putting aside the poor agreement
with the combined Homestake and GALLEX data, and to motivate a closer
inspection of the implications of this class of models.

In order to investigate this framework more completely, it must first be
embedded into a renormalizable model that can be applied up to the weak
scale. It is only then that it may be confronted with the complete range of
experimental information that is available to constrain potential neutrino
physics. We have made this comparison and find that the most stringent such
constraints come from three sources: (a) Nucleosynthesis, (b) Double Beta
Decay, and (c) Naturalness. We address each of these in turn.
(They are summarized in Table III.)

\midinsert
$$\vbox{\tabskip=0pt \offinterlineskip
\halign to \hsize{\strut#& #\tabskip 1em plus 2em minus .5em&
\hfil#\hfil &#& \hfil#\hfil &#\tabskip=0pt\cr
\noalign{\hrule}\noalign{\smallskip}\noalign{\hrule}\noalign{\medskip}
&& Criteron && Constraint &\cr
\noalign{\medskip}\noalign{\hrule}\noalign{\medskip}
&& Atmospheric, solar-$\nu$ anomalies
   && $\Avg{\phi_e}, \Avg{\phi_{e\mu}} \sim 1$ keV &\cr
&& && &\cr
&& Double beta decay && $\Avg{\phi_e}, \Avg{\phi_{e\mu}} \sim 1$ keV ,
			$M_{s_e} \roughly{>} 60 $ MeV &\cr
&&                   &&     &\cr
&& $\nu_e, \nu_\mu$ mixing with sterile $\nu$
   && $M_{s_e} \roughly{<} 10 $ MeV or $M_{s_e} \roughly{>} 350 $ MeV &\cr
&& && &\cr
&& Cosmology && $M_{s_e} \roughly{>} 2.3$ MeV ,
		0.7 MeV $\roughly{<} M_{s_\mu} \roughly{<} 2.3 $ MeV  &\cr
&& && &\cr
&& Naturalness && $M_{s_e}, M_{s_\mu} \roughly{<} 1 $ MeV &\cr
\noalign{\medskip}\noalign{\hrule}\noalign{\smallskip}\noalign{\hrule}
}}$$

\centerline{\bf Table III} \smallskip
\noindent {\eightrm A summary of the constraints which must
be satisfied by the class of models defined in section 2 and the
renormalizable model of section 4 in particular.  The lepton number
carrying scalars $\ss \phi_e,\phi_{e\mu}$ are defined in section 2.  The
lepton number carrying heavy sterile neutrinos $\ss S_e, S_\mu$ and their
masses $\ss M_{s_e}, M_{s_\mu}$ are defined in section 4.  Only the
supersymmetric generalization of the model in section 4 can
incorporate all the criterea listed in column one and it does so by
removing the naturalness constraint as given in the last row above.}
\endinsert

Any model for which neutrino decays play a role in explaining the
solar-neutrino deficit almost inevitably involves very light degrees
of freedom which couple appreciably to the electron neutrino. They are
therefore typically hard pressed to account for the success of the
standard Big-Bang framework of primordial nucleosynthesis, which
tightly constrains the number of relativistic degrees of freedom that
can be present when the temperature of the universe is of order (0.1
-- 1) MeV. A conservative attitude is to simply ignore this
constraint, with the rationale that the uncertainties involved in
predicting the behaviour of the universe at such an early epoch are
potentially quite large.  Although we are basically sympathetic to
this point of view, we have chosen here to see what is required to
accomodate this remarkable success of standard cosmology. We find that
consistency is possible, along the lines of the mechanism suggested in
Ref.~\negneutrino, provided that a relatively heavy
state (such as a sterile neutrino) exists with a mass around 1.5 MeV. In this
case we predict $N_{\rm eff} \simeq 2$, in good agreement with the data.

The double-beta decay excess events also point to a heavy neutrino state,
although in this case potentially much heavier than a few MeV. This condition
arises as a special case of a more general result which holds for
any renormalizable model for which all light scalars are electroweak singlets
\jim. The result holds because for these types of models no direct coupling is
possible between the electron neutrino and the light scalar that is emitted in
the decay. The act of emission therefore only occurs because of the mixing that
arises due to the mismatch between the neutrino electroweak and mass
eigenstates. The decay rate therefore goes to zero in the limit that all
neutrinos are degenerate, and so in particular when all neutrinos are light in
comparison with the MeV-scale energies that characterize the decay.

By contrast with the previous considerations which point toward the
existence of heavy neutrinos with masses in the MeV range, naturalness
considerations generically prefer all new particle states to be
lighter than this scale.  Otherwise loops in which these heavy
particles circulate produce contributions to the light scalar
potential that are much larger than those that are permitted if the
scalars are to have {\it v.e.v.}'s that are $O(1~\keV)$. In the
generic case, as may be seen from Table II and III, this naturalness
criterion conflicts with the requirements of nucleosynthesis and
double-beta decay, although only by one or two orders of magnitude.

If these naturalness considerations are eschewed, our renormalizable model
of Section (4) satisfactorily accounts for the solar and atmospheric
neutrino deficits, explains the excess double-beta decay events, and satisfies
all phenomenological constraints, including those from primordial
nucleosynthesis. It robustly predicts that future solar-neutrino experiments
will find some oscillations from electron- to muon-type neutrinos, together
with an overall depletion of the solar flux due to neutrino decays with a
lab-frame lifetime in the vicinity of 1000 sec. It also predicts that
lower-energy neutrinos are more depleted than are higher-energy ones, requiring
either the Homestake or the GALLEX signals to presently be incorrect at
the several $\sigma$ level.

In the particular renormalizable realization that is explored here, the model
also predicts the existence of sterile neutrinos in both the few MeV and the
several-hundred MeV ranges. The MeV neutrino is appreciably mixed with
$\nu_\mu$, and the hundred MeV neutrino mixes significantly with $\nu_e$. Both
types of mixings may be amenable to detection through closer scrutiny of the
$\pi \to \mu \nu$ decay spectrum, and in weak-universality measurements of the
electron charged-current interaction strength.

Taking the naturalness requirement seriously, one may simply accept very
light neutrino states, by ignoring the conflict with nucleosynthesis and the
double-beta decay anomaly. Alternatively, if the low-energy model is embedded
within a supersymmetric framework, all three types of constraints may be
satisfied. In particular, we have shown how supersymmetry can be consistent
with
weak-scale masses for the superpartners of all presently-known particles, as
well as keV-scale splittings among the new light supermultiplets. This
keV-scale
splitting is what sets the size of the light scalar potential, and ensures
the proper coupling strength for these scalars in the low-energy
phenomenological
model.

In any of these scenarios we find that the atmospheric, solar and double beta
decay anomalies lead us towards new scalars at very low energy scales, and new
neutrino physics at MeV scales.

\bigskip \centerline{\bf Acknowledgements} \bigskip

We would like to thank Hamid Bougourzi for collaborations on a related problem,
and
David Britton for going out of his way to provide us with the bounds on
heavy neutrino-lepton mixing. Thanks also to Fernando Quevedo and
Sandip Pakvasa for their helpful conversations.  This research was partially
funded by the N.S.E.R.C.\ of Canada and le Fonds F.C.A.R.\ du Qu\'ebec.

\vfill\eject

\section{Figure Captions}

\topic{Figure (1)}
One-loop contributions to the small scalar couplings which arise due to
couplings with the heavy Standard-Model scalar.
\topic{Figure (2)}
Dangerous one-loop contributions to the small scalar couplings due to loops
with heavy neutrinos.
\topic{Figure (3)}
The supergraphs which most strongly couple the SUSY-breaking scale of the SSM
to the light multiplets.
\endtopic

\listrefs

\bye